\documentclass[ ,final] {aipproc}

\layoutstyle{6x9}

\def\vom{\vec\Omega}

\begin{document}

\title{Spin tune in the single resonance model with a pair of 
Siberian Snakes}

\author{{\underline{D.~P.~Barber}}} {
  address={Deutsches~Elektronen--Synchrotron, DESY, ~22603 ~Hamburg,~Germany.}
}

\author{R. Jaganathan}{
  address={The Institute of Mathematical Sciences,   
Chennai. Tamilnadu  600113, India.
}
}

\author{M. Vogt} {
  address={Deutsches~Elektronen--Synchrotron, DESY, ~22603 ~Hamburg,~Germany.}
}

\begin{abstract}
Snake ``resonances'' are classified in terms of the invariant spin
field and the amplitude dependent spin tune.  Exactly at snake
``resonance'' there is no continuous invariant spin field at most
orbital amplitudes.
\end{abstract}

\maketitle


\section{Prologue}
This is an extended version of the paper with the same title 
published in the proceedings of the conference
SPIN2002 \cite{spin2002}.  A key aspect of the original paper was that
the invariant spin field for snake ``resonances'' is irreducibly
discontinuous at most orbital amplitudes. However,
details were omitted 
owing to the page limit.  In the
meantime other papers \cite{mane2004, beh2004} have appeared which
discuss the invariant spin field at snake ``resonances'' and it has
become clear that it would be useful to extend \cite{spin2002} to give more
details.

We begin by presenting a slightly polished version of the original paper.
Then the additional material 
is presented as an addendum. The citations are also updated.

\section{Introduction}

Spin motion in storage rings and circular accelerators is most
elegantly systematised in terms of the invariant spin field (ISF) and
the amplitude dependent spin tune (ADST). 
Here we apply them in the context of snake ''resonances''.
We begin by briefly recapitulating some necessary basic ideas. 
For more details see \cite{hh96,hvb99,gh2000,mv2000}. 

Spin motion in electric and magnetic fields at the
6--dimensional phase space point $\vec z$ and position $s$ around the ring,
is described by the T--BMT precession equation 
$d \vec{S}/{d s}= \vom (\vec z; s) \times \vec {S}$ ~\cite{jackson,hh96}
where $\vec S$ is the
spin expectation value (``the spin'') in the rest frame of the particle and $\vom (\vec z; s)$ 
contains the electric and magnetic fields in the laboratory.
The ISF, denoted by $\hat n(\vec z; s)$, is a 3--vector {\em field} of 
unit length obeying the T--BMT equation along particle 
orbits $(\vec z(s); s)$
and fulfilling the
periodicity condition  ${\hat n}(\vec z; s + C) = {\hat n}(\vec z; s)$
where $C$ is the circumference. Thus  
${\hat n}({\vec M}(\vec z; s); s + C) = {\hat n}({\vec M}(\vec z; s); s) =
R_{_{3 \times 3}}(\vec z; s) {\hat n}(\vec z; s)$ 
where ${\vec M}(\vec z; s)$ is the new phase space
vector after one turn starting at $\vec z$ and $s$ and 
$R_{_{3 \times 3}}(\vec z; s)$ is the corresponding spin transfer matrix.
The scalar product $J_{\rm s} = \vec S \cdot \hat n$ 
is invariant along an orbit,
since both vectors obey the T--BMT equation. Thus
with respect to the 
local $\hat n$ the 
motion of  $\vec S$  is 
simply 
a precession  around  $\hat n$.
The field $\hat n$ can be constructed at each 
reference energy where it exists without reference to individual spins.

The chief aspects of the ISF are that: 
1) For a turn--to--turn invariant particle distribution in phase space, 
 a distribution of spins initially aligned along the ISF remains invariant 
(in equilibrium) from turn--to--turn,~
2) for integrable orbital motion and away from orbital resonances the ISF  
determines the maximum attainable time averaged polarisation
$P_{_{\rm lim}} = |\langle\hat n(\vec z;s)\rangle|$
on a phase space torus at each $s$, 
where $\langle \rangle$ denotes the average  over the orbital phases,
3) under appropriate conditions $J_{\rm s}$ is an adiabatic invariant while 
system parameters such as the reference energy are slowly varied,
4) it provides the main axis for orthonormal coordinate systems
at each point in phase space which serve to define the ADST 
which in turn is used to define the concept of spin--orbit resonance.

These coordinate systems are constructed by attaching two other unit vectors
$\hat u_1(\vec z; s)$ and $\hat u_2(\vec z; s)$  to all ($\vec z, s$)
such that the sets ($\hat u_1, \hat u_2,\hat n$) are orthonormal. 
Like $\hat n$, the {\em fields} $\hat u_1$ and $\hat u_2$ are 1--turn periodic 
in $s$:~
${\hat u}_i(\vec z; s + C)={\hat u}_i(\vec z; s)$ for $i {\in} \{1,2\}$.
With the basis vectors $\hat u_1$ and $\hat u_2$ we can  
quantify the rate of the above mentioned spin precession around  ${\hat n}$: 
it is the rate of rotation of the  projection of $\vec S$ onto the
$\hat u_1, \hat u_2$ plane. 
Except on or close to orbital resonance, the  fields 
$\hat u_1(\vec z; s)$ and $\hat u_2(\vec z; s)$ can be chosen so that
the rate of precession is constant and independent of the 
orbital phases \cite{ky86,hvb99,gh2000,mv2000}.
The number of 
precessions per turn ``measured'' in this way is called the spin tune. 
The spin tune, $\nu_{{\rm s}}( \vec J)$, depends only on the orbital 
amplitudes (actions) $\vec J$, hence the name ADST.
The choice of some $\hat u_1(\vec z; s)$ and $\hat u_2(\vec z; s)$
satisfying the condition 
${\hat u}_i(\vec z; s + C)={\hat u}_i(\vec z; s)$ for $i {\in} \{1,2\}$
is not unique. An infinity of others can be chosen by suitable rotations
of the ${\hat u}_i$ around ${\hat n}$. 
These lead to the {\em equivalence class} of spin tunes obtained by the
transformation:
${\nu}_{{\rm s}}(\vec J) \Rightarrow 
{\nu}_{{\rm s}}(\vec J) + l_0 + l_1 Q_1 + l_2 Q_2 + l_3 Q_3$
for any integers $l$ where the $Q(\vec J)$'s are the tunes on a torus of 
integrable orbital motion 
\footnote{For a recent detailed discussion of these concepts, see \cite{beh2004}.}.
The ADST provides a way to quantify the degree of coherence between
the spin and orbital motion and thereby predict how strongly the electric 
and magnetic fields along particle orbits disturb spins.
In particular, the spin motion can become very erratic close to the
{\em spin--orbit resonance} condition
$\nu_{\rm s} (\vec J) ~=~m_0 + m_{1} Q_{1} + m_{2} Q_{2} + m_{3} Q_{3}$
where the $m$'s are integers.
At these resonances the ISF can spread out 
so that $P_{_{\rm lim}}$ is very small. Examples of the behaviour  
of $P_{_{\rm lim}}$ near 
spin--orbit resonance and the application of a generalised
Froissart--Stora description of the breaking of the adiabatic invariance
of $J_{\rm s}$ while crossing resonances during variation of system parameters 
can be found in \cite{gh2000,mv2000,spin2000,hv2004}.
Note that: 1)~the resonance condition is {\em not} expressed in 
terms of the spin tune $\nu (\vec 0)$ on the closed orbit, 2)~a ``tune'' 
describing spin motion but depending on orbital phases could not
be meaningful in the spin--orbit resonance condition, 3)~if the
system is on spin--orbit resonance for one spin tune of the 
equivalence class, it is on resonance for all others.
In general $\hat u_1$ and $\hat u_2$ do not obey the T--BMT 
equation along an orbit $(\vec z(s); s)$. 
But at spin--orbit resonance, they  
can be chosen so that a spin $\vec S$ is at rest in its 
local ($\hat u_1, \hat u_2,\hat n$) system. 
Then $\hat u_1(\vec z; s)$ and $\hat u_2(\vec z; s)$ do obey the T--BMT 
equation so that the ISF $\hat n(\vec z; s)$ is not unique. 

Nowadays we emphasise the utility of the ISF for 
defining equilibrium spin distributions. However, it was originally
introduced for bringing the combined semiclassical Hamiltonian
of spin--orbit motion into action--angle form for calculating the effects 
of synchrotron radiation \cite{mont98}.  
The initial Hamiltonian is written as
$H_{\rm s-o} = 
\frac{2 \pi}{C} ( Q_1 J_1 + Q_2 J_2 + Q_3 J_3) + \vom \cdot \vec S$.
By viewing the  spin motion in the ($\hat u_1, \hat u_2,\hat n$)
systems, a new Hamiltonian in full action--angle form
$H^{\rm aa}_{\rm s-o} = 
\frac{2 \pi}{C} ( Q'_1 J'_1 + Q'_2 J'_2 + Q'_3 J'_3) + \frac{2 \pi}{C} 
{\nu}_{{\rm s}}({\vec J}') J_{\rm s}$ is obtained which is valid at first 
order in $\hbar$ \cite{ky86}. 
This  emphasises again that, as with all 
action--angle formulations, the  spin frequency cannot depend on orbital 
phases. Moreover, it is easy to show that at orbital resonance, ~( i.e.~
$~k_0 + k_{1} Q_{1} + k_{2} Q_{2} + k_{3} Q_{3} = 0$ for suitable integers
$k$) the ``diagonalisation'' of the Hamiltonian (i.e. finding the 
$\hat u_1,\hat u_2$) might not be possible \cite{ky86}. 
Thus at orbital resonance
the ADST may not exist. On the other hand, one avoids running a machine
on such resonances. The spin tune on the closed orbit 
$\nu_0 = \nu_{{\rm s}}( \vec 0)$ always exists and so does 
$\hat n_0 (s) = \hat n(\vec 0; s)$.

For our present purposes there are two kinds of orbital resonances: 
resonances where at least one of the $Q$'s is irrational and those
where all are rational. We write the rational tunes as 
$Q_i = a_i/b_i$ ($i = 1,2,3$) where the $a_i$ and $b_i$ are integers.
Then for the second type, the orbit is periodic over $c$ turns where $c$  
is the lowest common multiple of the $b_i$. This opens the possibility that in this case the ISF at each $(\vec z, s)$ can be obtained (up to a sign) 
as the unit
length real eigenvector of the $c$--turn spin map (c.f. the calculation
of $\hat n_0$ from the  1--turn spin map on the closed orbit.). 
However, the corresponding eigentune $c \nu_{\rm c}$ extracted from the 
complex eigenvalues $\lambda_{\rm c} = e^{\pm 2\pi i c \nu_{\rm c}}$,
depends in general on the orbital 
phases at the starting $\vec z$. Thus in general $\nu_{\rm c}$ is {\em not}
a spin tune and should not be so named \cite{mane}. 
Nevertheless if  $c$ is very large the dependence of $\nu_{\rm c}$ on the 
phases 
can be very weak so that it can approximate well the ADST of nearby 
irrational tunes. This is expected heuristically since the 
influence of the starting phase can be diluted on forming the spin map for 
a large number of turns. 
At non--zero amplitudes, both for irrational or rational $Q$'s, the
eigentune of the 1--turn 
spin map usually has no physical significance. Of course, it normally depends on the 
orbital phases and the corresponding eigenvectors are normally not even solutions
of the T--BMT equation.   

For non--resonant orbital tunes, the spin tune can be obtained 
using the
SODOM--II algorithm \cite{ky99} whereby spin motion is written in terms of 
two component spinors and SU(2) spin transfer matrices. 
The {\em functional equation}
${\hat n}({\vec M}(\vec z; s); s) =
R_{_{3 \times 3}}(\vec z; s) {\hat n}(\vec z; s)$
is then expressed in terms of a
Fourier representation, w.r.t. the orbital phases,
of the spinors and of the 1--turn SU(2) matrices.
The spin tune appears as the set of eigentunes of an {\em eigen problem for
Fourier components}  and $\hat n$ is reconstructed from
the Fourier eigenvectors.
SODOM--II delivers the whole spin field on the torus
$\vec J$ at the chosen $s$.

\section{The single resonance model}

In perfectly aligned flat rings with no solenoids, $\hat n_0$ is vertical
and $\nu_0$ is in the equivalence class containing $a \gamma_0$ where 
$\gamma_0$ is the Lorentz
factor on the closed orbit  and $a$ is the gyromagnetic anomaly of the particle. 
In the absence of skew quadrupoles, the primary disturbance to spin
is then from the radial magnetic fields along vertical betatron trajectories.
The disturbance can be very strong and the polarisation
can fall if the
particles are accelerated through the condition 
$a \gamma_0 =  \kappa \equiv k_0 \pm Q_2$ where mode 2 is vertical motion.
This can be understood in terms of the ``single resonance model''
(SRM) whereby a rotating wave  approximation is made in which 
the contribution to $\vom$ from  
the radial fields along the orbit is 
dominated by the Fourier harmonic at $\kappa$  
with strength $\epsilon (J_2)$.
The SRM  can be solved exactly and 
the ISF is given by \cite{mane2} 
${\hat n}(\phi_2) = {\rm sgn}(\delta) 
\left( \delta {\hat e}_2 +  
\epsilon ({\hat e}_1 \cos \phi_2 + {\hat e}_3 \sin \phi_2 ) \right)/
\sqrt {\delta^2 + \epsilon^2}$
where $\delta = a \gamma_0 - \kappa$, $\phi_2$ is the orbital phase,
$({\hat e}_1,{\hat e}_2,{\hat e}_3)$ are horizontal, vertical and 
longitudinal unit vectors
and the convention ${\hat n \cdot {\hat e}_2} \ge 0$ is used.
The tilt of $\hat n$ away from the vertical $\hat n_0$ is 
$| \arcsin (\epsilon /\sqrt {\delta^2 + \epsilon^2}) |$ 
so that it is $90^{\circ}$ at $\delta = 0$ for non--zero $\epsilon$.
At large $|\delta |$, the equilibrium 
polarisation  directions $\hat n(J_2,\phi_2; s)$, are almost parallel 
to $\hat n_0(s)$  but during acceleration through
$\delta = 0$, $\hat n$ varies strongly and the polarisation  
will change if the adiabatic  invariance of  $J_{\rm s}$ violated. 
The change in $J_{\rm s}$ for acceleration through $\delta = 0$
is given by the Froissart--Stora formula.
The  ADST which reduces to $a \gamma_0$ on the closed orbit is 
$\nu_{\rm s} = {\rm sgn}(\delta){\sqrt {\delta^2 + \epsilon^2}} + 
\kappa$. 
Note that the condition $\delta = 0$ is {\em not} the spin--orbit
resonance condition. On the contrary, as $\delta$ passes through zero
$\nu_{\rm s}$ jumps by $2 \epsilon$ with our convention 
for $\hat n$ and avoids fulfilling the true
resonance condition: for particles with non--zero $\epsilon$,  $a \gamma_0$
is just a parameter.
In this simple model $\nu_{\rm s}$ exists and is well defined near 
spin--orbit resonances for all $Q_2$. 
This is also true in more general cases 
if orbital resonance is avoided.

\section{The single resonance model with a pair of Siberian Snakes}
\subsection{Snake ``resonances''}
Polarisation loss while accelerating through $\delta = 0$ can be 
reduced by installing
pairs of Siberian Snakes, magnet systems
which rotate spins by $\pi$ independently of $\vec z$ around 
a ``snake axis'' in the machine plane. 
For example, one puts two snakes at diametrically opposite points on the ring.
Then 
${\hat n_0 \cdot {\hat e}_2} = +1$ in one half ring and $-1$ in the other.
With the snake axes relatively at $90^{\circ}$, $\nu_0$ is in the equivalence 
class containing $1/2$ for all $\gamma_0$.
For calculations one often represents the snakes as elements of zero length
(``pointlike snakes''). Then if, in addition, the effect of 
vertical betatron motion is described by the 
SRM, and orbital resonances are avoided, 
calculations with SODOM--II, perturbation theory \cite{ky88} and 
the treatment in \cite{mane} suggest that  
$\nu_{\rm s}(J_2)$ is in the equivalence class containing $1/2$ too, 
independently of $\gamma_0$ but also of
$J_2$. Thus for $Q_2$ sufficiently away from $1/2$ no 
spin--orbit resonances 
$\nu_{\rm s}(J_2) = k_0 \pm Q_2$ are crossed during acceleration 
through $\delta = 0$ and the polarisation can be preserved. 
This is confirmed by 
tracking calculations. However, such calculations and 
analytical work show that the polarisation can still be lost 
if the fractional part of $Q_2$, 
$[Q_2]$, is ${\tilde a}_2/2 {\tilde b}_2$  where here, and later,
${\tilde a}_2$ and ${\tilde b}_2$ are odd positive integers with 
${\tilde a}_2 < 2 {\tilde b}_2$ and where here and later the brackets $[...]$ 
are used to signal the fractional part of a number. 
This is the so--called ``snake resonance phenomenon'' and it also  has 
practical consequences \cite{syl,luccio,mbai2004}, especially for small ${\tilde b}_2$.
Such a $[Q_2]$ fits the condition 
$1/2 =  (1 - {\tilde a}_2)/2 + {\tilde b}_2 [Q_2]$. 
But calculations (see below) show that exactly at
$[Q_2] = {\tilde a}_2/2 {\tilde b}_2$ the ADST may not exist. 
If it doesn't, it isn't in the 
equivalence class for $1/2$.  Then we are not dealing with a 
conventional
resonance $\nu_{\rm s}(J_2) =  (1 - {\tilde a}_2)/2 + {\tilde b}_2 [Q_2]$
and the term {\em resonance} is inappropriate.
Depolarisation in this model 
has also been attributed to the fact that for 
non--zero $J_2$ the 
eigentune of the 1--turn spin map, which depends on $\phi_2$, is 1/2 at some values of 
$\phi_2$ \cite{syl}. However,  such a quantity  
does not describe spin--orbit coherence.
Snake ``resonances'' are usually associated with
acceleration but it has been helpful in other circumstances 
\cite{gh2000,mv2000,spin2000} 
to begin by studying the {\em static} properties of the system, namely
with the ISF. We now do that for the SRM with two snakes for 
representative, parameters. 

\subsection{Numerical study}
Figure 1 shows $P_{_{\rm lim}}$ (just before a snake) and 
$\nu_{\rm s}$ 
for  25000 equally spaced $[Q_2]$'s between 0 and 0.5  for
$\epsilon = 0.4$ and $\delta = 0$. 
At each $[Q_2]$, $\hat n$ is calculated by stroboscopic averaging \cite{hh96}
($ \le25~10^6$ turns)
at 500 equally spaced $\phi_2$ in the range
$0 - 2\pi$ and $P_{_{\rm lim}}$ is obtained by averaging over these $\phi_2$.
The ADST is obtained from SODOM--II. If the ADST exists SODOM--II delivers
a part of the equivalence class, namely 
the spectrum $[\pm 0.5 +  l_2 Q_2]$ for a range of contiguous even $l_2$
restricted by the necessarily finite size of the matrix of Fourier 
coefficients. Only even $l_2$ are allowed by the algorithm. For irrational 
$Q_2$ the range of $l_2$ is large. For rational $Q_2$ the spectrum can
include $\pm 0.5$ but is otherwise highly degenerate or 
contains none or just a very few of the required members 
$[\pm 0.5 + l_2 Q_2]$. 
Thus the existence of an ADST is easily checked.
The central horizontal row of points in figure 1 shows 
the common  member $+0.5$ of the equivalence class of the ADST 
at the values of $[Q_2]$ where the ADST exists.
There is an ADST for most $[Q_2]$'s used.
The first row of dots up from the bottom marks $[Q_2]$ values
where there is no ADST. As expected,
these are all at rational $[Q_2]$'s such as 1/5, 1/4, 2/5 $\dots $
or ${\tilde a}_2/2 {\tilde b}_2 = 1/6, 3/14, 3/10 \dots$ and  the 
$[c \nu_{\rm c}]$ computed for these $[Q_2]$ show $\phi_2$   dependence.
The curved line shows $P_{_{\rm lim}}$ and the second row of dots from the
bottom marks $[Q_2]$ values where the ISF obtained by stroboscopic averaging
did not
converge for all phases.  These coincide with sharp dips in 
$P_{_{\rm lim}}$ and are at or near $[Q_2] = {\tilde a}_2/2 {\tilde b}_2$, 
i.e in the snake ``resonance''  subset of the $[Q_2]$'s  in the first row.
Thus  snake ``resonance'' is already a static phenomenon.
Near such $[Q_2]$'s, the ISF, which for just one orbital mode is a closed 
curve in three dimensions, becomes extremely complicated as $\hat n$ strives 
to  satisfy its defining conditions. Right
at  $[Q_2] = {\tilde a}_2/2 {\tilde b}_2$ the nonconvergence occurs at 
$[\phi_2/{2 \pi}] = j/{2 \tilde b}_2$ for
integers $j = 1,..., 2 {\tilde b}_2 $ and, moreover,
{\em the ISF is discontinuous at these phases}
\footnote{Note that in \cite{beh2004} it was
convenient to require that the magnetic and electric fields 
and the ISF were smooth in $\phi$ and $s$.  Here we drop that
requirement since we are dealing with models with pointlike
snakes.}.
For $[Q_2] = {\tilde a}_2/4 {\tilde b}_2$  (${\tilde a}_2 < 4 {\tilde b}_2$), 
$P_{_{\rm lim}}$ and the ISF
show no special behaviour. 
These observations are consistent 
with the perturbative result \cite{mv2000} that for mid--plane symmetric 
systems,
$\hat n$ should be well behaved near even $m_2$ 
but may show exotic behaviour close to odd $m_2 = {\tilde b}_2$. 
As expected, $P_{_{\rm lim}}$ and the ISF also show no special behaviour for
$[Q_2] = {a}_2/{\tilde b}_2$ (${a}_2 < {\tilde b}_2$).
Some snake ``resonances'' such as that at $[Q_2] = 1/30$ are 
narrower than 0.00002 in $[Q_2]$ and are missed in this scan. 
$P_{_{\rm lim}}$ also has several dips 
at values of $[Q_2]$ (e.g. at 0.341) which appear to have no special 
significance, but which  should still be avoided at storage. 
The results for $0.5 \le [Q_2] \le 1.0$ are the reflection in 0.5
of the curves and points shown. 
Qualitatively similar results are obtained with equally distributed
odd pairs of snakes set to give $\nu_0 = 1/2$.
The ISF and $P_{_{\rm lim}}$ usually vary 
significantly with $s$.
\begin{figure}[ht]
\includegraphics[width=15.0cm,height=5.9cm]{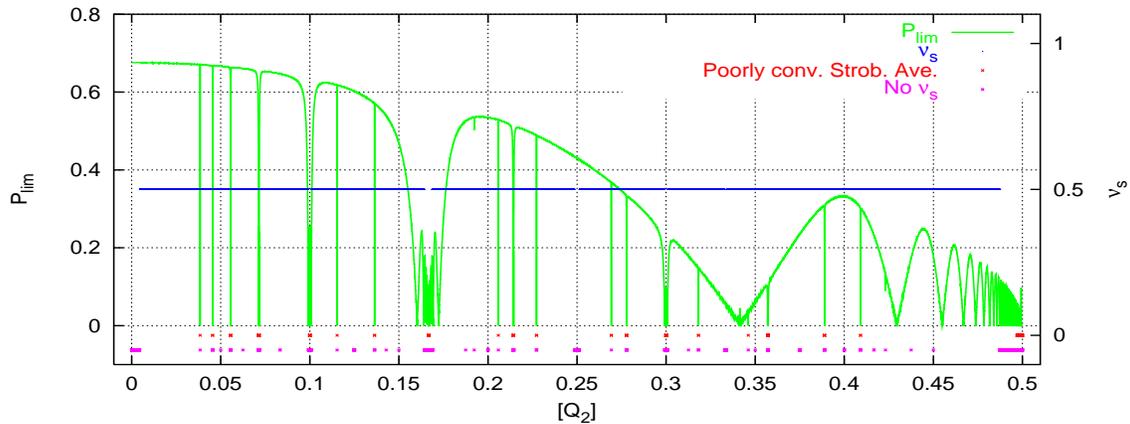}  
\caption{$P_{_{\rm lim}}$ (left axis) and a component of the ADST (right axis)
for the SRM with $\delta = 0$, $\epsilon = 0.4 $ and with 2 Siberian Snakes 
with axes at $90^{\circ}$ and $0^{\circ}$.
}
\end{figure}

\begin{figure}[htbp]
\includegraphics[width=12.0cm]{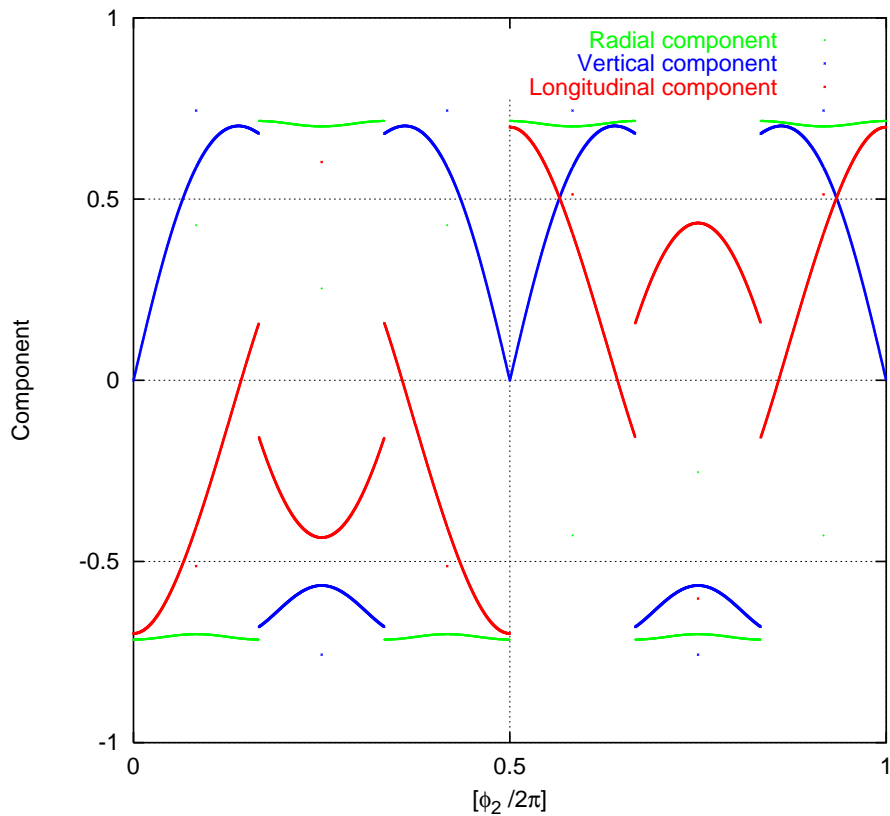}  
\caption{The three components of ${\hat n}(\phi_2)$ 
for the SRM with  2 Siberian Snakes 
with axes at $90^{\circ}$ and $0^{\circ}$ and for $[Q_2] = 1/6$. Viewing point: just before a 
snake. $\delta = 0$ and  $\epsilon = 0.4$.
}
\end{figure}

\subsection{Addendum}

A typical example for an ISF at a snake ``resonance'' is shown in
figure 2. This shows the components of the ISF for $[Q_2] = 1/6$ and
$\epsilon = 0.4$.  In this case they are obtained by using the 6--turn spin map to
calculate the vector $\hat n$ in the range $0 < [\phi_2/{2 \pi}] \le 
1/{2 \tilde b}_2$, namely 0 to 60 degrees, while applying the
constraint that $\hat n$ should be continuous in $\phi_2$. Then the
$\hat n$ for each $\phi_2$ in this range is transported with the
1--turn spin map for six, or more, turns. 
One sees that $\hat n$ changes sign at values of $\phi_2$ which are multiples of 60 degrees
so that the ISF is discontinuous as advertised. 
The six sets of stray points at $[\phi_2/{2 \pi}] = 1/12, 1/4 \dots$ 
are at  phases $\phi_2$ where the 6--turn spin map is the identity. At 
these phases $\hat n$ obtained in this way is arbitrary and the algorithm delivers 
values dominated by numerical noise.
Of course, if the ISF is represented as the locus of points on the
surface of the unit 2--sphere, this ISF gives disjoint segments.
It is also easy to demonstrate that the positions of the discontinuities can be shifted. 
Thus at snake ``resonances'' the invariant spin field is not only discontinuous but also {\em non--unique}
and to an extent which goes far beyond the non--uniqueness at the phases $\phi_2$ at which 
the 6--turn spin map is the identity.  The ISF shows analogous behaviour at other snake ``resonances''.
This is in stark contrast to the case, say, of the pure SRM.  
Note that discontinuities would not be allowed for irrational $[Q_2]$, but that they are
not prohibited for rational $[Q_2]$. Moreover,  theorems  on uniqueness of the 
ISF require the existence of 
a spin tune and that the system is away from orbital resonance.
Of course, for snake ``resonances'' with very high values of ${\tilde b}_2$, 
the ISF and the corresponding configuration of equilibrium polarisation is very complicated. 
It is then far from clear whether the ISF is a useful concept for these simple models involving 
just one plane of orbital motion and
singular, i.e., non--physical fields, although, 
of course, such models have been instrumental in presaging the loss of polarisation
observed in real storage rings \cite{syl,luccio, mbai2004}. 

The ISF obtained from the real eigenvector of the multi--turn spin map is also non--unique for $[Q_2] = {\tilde a}_2/4 {\tilde b}_2$, namely at those $[\phi_2/{2 \pi}]$ where the $4 {\tilde b}_2$--turn spin map is the identity.
Moreover, additional discontinuities in the sign can be added by hand, thereby enhancing the choice of ISF's. But in contrast to the case of 
snake ``resonances'' such discontinuities remain optional.

Further aspects of these matters will be reported elsewhere.

It is instructive to compare the discontinuous curves in figure 2 with the smooth curves in  figures 7 and 8 in
\cite{mane2004}
\footnote{It is also shown in \cite{mane2004} that the curve in figure 1 can be reproduced 
by the MILES algorithm by the same author. MILES, like SODOM--II, is based on SU(2) and Fourier expansions.}. 
These are also said to represent ISF's at snake ``resonances''.
The vectors corresponding to the curves in figure 2
satisfy the T--BMT equation by construction and they are single valued
in $[\phi_2/{2 \pi}]$ as required for an ISF.  However, if the vectors
for the curves in figures 7 and 8 in \cite{mane2004} are transported
according to the T--BMT equation, they are not single valued in
$[\phi_2/{2 \pi}]$. Alternatively, if those curves are taken to
represent single--valued functions of $[\phi_2/{2 \pi]}$, as depicted,
then they do not represent spin motion, i.e., motion according to the T--BMT equation.
Either way, the curves in those figures do not represent ISF's at snake ``resonances''.

\section{Summary}
A snake ``resonance'' is at root a {\em static} phenomenon
characterised by an invariant spin field
which, for the simple models discussed here,  is irreducibly discontinuous in $\phi_2$ for most orbital amplitudes.
Moreover, on and near snake ``resonance'', there is no 
amplitude dependent spin tune so that 
the snake ``resonances'' of these models are not 
simple spin--orbit resonances. 
The mechanism, in terms of $J_{\rm s}$, for polarisation loss  during 
acceleration through $\delta = 0$ at and near such $[Q_2]$'s is under study. \\

We thank K. Heinemann, G. H. Hoffstaetter and  J.A. Ellison for useful discussions.

\end{document}